# Growth and properties of ferromagnetic In$_{1-x}$Mn$_x$Sb alloys

T. Wojtowicz[a,b,*], W.L. Lim[a], X. Liu[a], G. Cywiński[b], M. Kutrowski[a,b], L. V. Titova[a], K. Yee[a], M. Dobrowolska[a], J. K. Furdyna[a], K. M. Yu[c], W. Walukiewicz[c], G.B. Kim[d], M. Cheon[d], X. Chen[d], S.M. Wang[d], H. Luo[d], I. Vurgaftman[e], J.R. Meyer[e]

[a] *Department of Physics, University of Notre Dame, Notre Dame, IN 46556*
[b] *Institute of Physics, Polish Academy of Sciences, 02-668 Warsaw, Poland*
[c] *Electronic Materials Program, Materials Sciences Division, Lawrence Berkeley National Laboratory, Berkeley, CA 94720*
[d] *Department of Physics University at Buffalo, State University of New York, Buffalo, NY 14260*
[e] *Code 5613, Naval Research Laboratory, Washington, DC 20375*

ABSTRACT

We discuss a new narrow-gap ferromagnetic (FM) semiconductor alloy, In$_{1-x}$Mn$_x$Sb, and its growth by low-temperature molecular-beam epitaxy. The magnetic properties were investigated by direct magnetization measurements, electrical transport, magnetic circular dichroism, and the magneto-optical Kerr effect. These data clearly indicate that In$_{1-x}$Mn$_x$Sb possesses all the attributes of a system with carrier-mediated FM interactions, including well-defined hysteresis loops, a cusp in the temperature dependence of the resistivity, strong negative magnetoresistance, and a large anomalous Hall effect. The Curie temperatures in samples investigated thus far range up to 8.5 K, which are consistent with a mean-field-theory simulation of the carrier-induced ferromagnetism based on the 8-band effective band-orbital method.

---

[*]Author to whom correspondence should be addressed; electronic mail: wojto@ifpan.edu.pl

## 1. Introduction

The recent emergence of III-Mn-V semiconductors, whose unique properties include the potential for externally-controlled ferromagnetism, has led to a number of innovative application concepts [1-4]. Non-cryogenic "spintronic" device operation is desired ultimately, hence the research activity has focused largely on those alloys with the smallest lattice parameters and largest energy gaps (such as $Ga_{1-x}Mn_xAs$, and more recently $Ga_{1-x}Mn_xN$ and $In_{1-x}Mn_xN$ [5]), which are expected to yield the highest Curie temperatures $T_C$ [2,6]. However, many aspects of the III-Mn-Vs remain unknown, and even the fundamental mechanism for ferromagnetism is understood only qualitatively [5,7]. In this context, we believe that a great deal can be learned by exploring the opposite extreme of the family, $In_{1-x}Mn_xSb$, which has the *largest* lattice constant and *smallest* energy gap. From the applications standpoint, the small energy gap of $In_{1-x}Mn_xSb$ offers prospects for infrared spin-photonics. Carrier transport should also be enhanced, due to a smaller hole effective mass and higher hole mobility.

Non-ferromagnetic Mn-doped InSb was grown previously in bulk as a possible ultra-low-temperature thermoresistor material [8], and by molecular beam epitaxy (MBE) for magnetic field sensing [9]. The most closely-related prior ferromagnetism study used metal organic MBE to produce the InMnAs-rich quaternary $In_{1-x}Mn_xAs_{0.8}Sb_{0.2}$, whose magnetization displayed interesting light-induced changes [10].

## 2. Experimental procedures

The physical properties of the $In_{1-x}Mn_xSb$ epilayers were studied by x-ray diffraction (XRD), SQUID magnetometry, electrical transport (magnetoresistance and anomalous Hall effect), magnetic circular dichroism (MCD), magnetooptical Kerr effect (MOKE), channeling Rutherford backscattering (c-RBS), and channeling particle-induced x-ray emission (c-PIXE).

[11]. The films were grown in a Riber 32 R&D MBE system. Fluxes of In, Be, and Mn were supplied by standard effusion cells, while $Sb_2$ was produced by an Sb cracker cell. To avoid the large parallel conductance that would result if a thick InSb buffer layer were grown at high temperature, our $In_{1-x}Mn_xSb$ films were deposited on closely-lattice-matched (001) hybrid CdTe/ZnTe/GaAs substrates, which were pre-fabricated by MBE at the Institute of Physics of the Polish Academy of Sciences. 2 nm thick ZnTe layer in hybrid substrates served to stabilize the (001) orientation of 4.5 μm thick CdTe layer. The growth of CdTe was performed at 350 ºC from 7N Cd and Te elemental sources, and with Cd:Te beam equivalent pressure ratios of 3:2. The CdTe layer was etched for a few seconds in 0.25% $Br_2$/methanol solution before the substrate was inserted into the III-V MBE chamber. This produced an oxide-free surface, whose protective top layer of Te was easily removed by heat within the growth chamber. We then grew a 100-nm low-temperature (LT) InSb buffer layer at 210 ºC, which provided a flat surface for deposition of the ferromagnetic alloy. The substrate was then cooled further, to 170 ºC, for LT growth of the $In_{1-x}Mn_xSb$, $In_{1-x-y}Mn_xBe_ySb$ or $In_{1-y}Be_ySb$ film to a thickness of ≈ 230 nm. All of these growths used an $Sb_2$:In beam equivalent pressure ratios of 3:1, and alloy concentrations were controlled by the Mn and Be effusion cell temperatures. Streaky RHEED patterns observed during the LT deposition of all three alloys indicated high-quality two-dimensional growth with a 1×3 reconstruction. RHEED intensity oscillations provided a measure of the growth rate (typically around 0.26 ml/s), and hence the thickness of each layer.

3. Experimental results and discussion

The crystallographic quality of the $In_{1-x}Mn_xSb$/InSb/CdTe layers and the lattice constants ($d_0$) of the $In_{1-x}Mn_xSb$ films were determined by XRD. Using the InSb elastic constants to approximate those of In$_{1-x}$Mn$_x$Sb and assuming that the InSb layer on CdTe was fully relaxed (since the initial growth of InSb was three-dimensional, $d_0$ = 0.64794 nm) while the In$_{1-x}$Mn$_x$Sb layer on InSb was fully strained (due to the low growth temperature), we found that the lattice constant for In$_{1-x}$Mn$_x$Sb decreased to 0.64764 (0.64762) nm for layers grown with a Mn cell temperature of $T_{Mn}$ = 710 (720) °C. From the c-RBS/PIXE data, the two samples had compositions $x$ = 0.02 (0.028). The decrease in lattice parameter is a consequence of the smaller covalent radius of Mn as compared to the In atom it replaces [12]. X-ray rocking curves for the In$_{1-x}$Mn$_x$Sb/CdTe specimens showed full widths at half maximum (FWHM) of ≈ 200 arc-sec (as compared to 160 arc-sec for the CdTe peak), indicating that the magnetic alloy had good crystal quality. Finally, for the range of compositions discussed here, the In$_{1-x}$Mn$_x$Sb x-ray data showed no indication that other crystallographic phases (*e.g.*, MnSb) were present.

Our recent studies of Ga$_{1-x}$Mn$_x$As [11] demonstrated that the ferromagnetic properties of a III-Mn-V semiconductor depend critically on the distribution of Mn atoms within the III-V host lattice, since Mn atoms in different locations play quite different roles. The ferromagnetism relies on both the concentration of Mn$^{++}$ 5/2 spins and the density of free holes that mediate their magnetic interaction through *p-d* coupling. However, Mn atoms can occupy three distinct lattice sites within the III-V host: (1) substitutional group-III positions (Mn$_{III}$), where as acceptors they supply both free holes and spins, (2) interstitial positions (Mn$_I$), where they act as compensating double donors, and (3) random positions (Mn$_R$), that are incompatible with the zinc-blende structure (*e.g.*, in a MnV precipitate) and are outside the picture of carrier-mediated ferromagnetism. It has been shown theoretically that interstitial Mn ions do not contribute to the hole-mediated ferromagnetism [13,14], because of negligible *p-d* coupling to the holes [15]. Moreover, a strong antiferromagnetic coupling [15] between Mn$_I$-Mn$_{III}$ pairs brought together by electrostatic attraction [11] further reduces the effective concentration of ferromagnetically-active $Mn_{III}$ ions, to the value $x_{eff} = (r_{III} - r_I)x$. Here $r_{III}$ and $r_I$ are, respectively, the atomic fractions of substitutional and interstitial Mn.

In order to determine the total Mn concentration and the location of Mn within the host lattice, we performed simultaneous channeling particle induced x-ray emission and channeling Rutherford backscattering measurements using a 1.95MeV $^4$He$^+$ beam (for a description of the technique see, *e.g.*, [11]). Mn K$_\alpha$ x-ray signals obtained by c-PIXE were compared directly with InSb c-RBS signals from the In$_{1-x}$Mn$_x$Sb films. From a quantitative analysis of the c-RBS/PIXE results, we obtain $x = 0.02$, $r_{III} = 0.85$, $r_I = 0.05$, $r_R = 0.1$ for the sample grown with $T_{Mn} = 710$ ºC and $x = 0.028$, $r_{III} = 0.82$, $r_I = 0.08$, $r_R = 0.1$ for the sample grown with $T_{Mn} = 720$ ºC. Here $r_{III}$ and $r_I$ are the fraction of Mn ions occupying group-III sites (In in our case) and interstitial sites, respectively, as defined above, while $r_R$ is the fraction of Mn ions on random sites that do not participate in the ferromagnetism. In agreement with our previous studies of Ga$_{1-x}$Mn$_x$As [11], the fraction of interstitial Mn ions increases strongly with total Mn concentration. A substantial Mn$_I$ concentration is, however, observed even in In$_{1-x}$Mn$_x$Sb with relatively low $x$. We expect the maximum hole concentration (limited thermodynamically through the Fermi-level-dependent defect formation energy [16]) that can be achieved in In$_{1-x}$Mn$_x$Sb for our growth conditions to be smaller than that in Ga$_{1-x}$Mn$_x$As. The electrical transport studies on In$_{1-x-y}$Mn$_x$Be$_{1-y}$Sb layers, discussed below, confirm that expectation.

Systematic magnetotransport studies were carried out on all of the specimens described in this paper, using a six-probe Hall bar geometry and ohmic indium contacts. Figure 1(a) shows the temperature dependence of the zero-field resistivity $\rho$ for In$_{1-x}$Mn$_x$Sb films with several nominal values of $x$ (as determined by the temperature of the Mn effusion cell). Figure 1(b) shows $\rho$ vs. $T$ for a series of $In_{1-x-y}Mn_xBe_{1-y}Sb$ layers grown with $T_{Mn}$ = 705 ºC and various Be contents $y$ (determined by the temperature of the Be effusion cell $T_{Be}$), which were investigated to examine the effects of $p$-type co-doping. Data for a non-magnetic $In_{1-y}Be_ySb$ layer with a similar hole concentration $p$ is also included in Fig. 1(b) for comparison. The value of $p$ in the $In_{1-y}Be_ySb$ layer was determined by Hall measurements to be $1.4 \times 10^{20}$ cm$^{-3}$. Note that the resistivities of the three $In_{1-x}Mn_xSb$ layers from Fig. 1(a), grown with $T_{Mn}$ = 700, 710 and 720 ºC, show clear maxima at $T_\rho$ = 5.4, 7.5 and 9.2 K, respectively. It is known from studies of $Ga_{1-x}Mn_xAs$ that these maxima signal the paramagnetic-to-ferromagnetic phase transition, and that $T_\rho$ occurs close to the Curie temperature $T_C$ [1,13]. Since $Ga_{1-x}Mn_xAs$ had previously been the only III-Mn-V semiconductor to display such distinct maxima, electrical transport studies of $In_{1-x}Mn_xSb$ provide a valuable new test case for theories of critical scattering at and near $T_C$. Figure 1(a) shows that the Curie temperature of $In_{1-x}Mn_xSb$ increases with increasing $x$ in the range studied here, as expected.

On the other hand, Fig. 1(b) illustrates that $T_\rho$ progressively *decreases* in layers co-doped with Be when the Be content $y$ ($T_{Be}$ varies from 0 to 930 ºC) is increased for fixed Mn concentration $x$. This finding is somewhat counter-intuitive, since it might be expected that adding Be to increase the hole concentration should enhance $T_C$ [2]. However, the observed drop in $T_C$ is consistent with our previous results for $Ga_{1-x}Mn_xAs$ samples co-doped with Be, and can be understood in terms of the same model [13,14,17]. When the free hole concentration associated with the substitutional Mn acceptors approaches the limit imposed by the formation energy for negatively-charge defects, which depends on the Fermi energy ($E_F$) [16], adding more Be acceptors to increase $E_F$ unavoidably induces the creation of compensating donors. That occurs primarily via a shift of Mn atoms from substitutional (acceptor) to interstitial (double-donor) sites [13,14,17]. Not only are the $Mn_I$ ions ferromagnetically inactive, but they form anti-ferromagnetic $Mn_{III}$-$Mn_I$ pairs as was pointed out above. The net result is a reduction in the density of active Mn spins, which in turn produces the observed drop of $T_C$.

Attribution of the $\rho(T)$ maximum to critical scattering is further corroborated by the observation of negative magnetoresistance, which occurs because the critical scattering decreases when an external magnetic field aligns the magnetic moments. This effect has in fact been observed for all of the III-Mn-V ferromagnetic semiconductors studied so far [1], and also for non-ferromagnetic bulk InSb doped with Mn [8,18]. The top panel of Fig. 2(a) shows that the magnetoresistance of $In_{0.98}Mn_{0.02}Sb$ is negative for all temperatures below 30 K. The amplitude of the magnetoresistance increases with decreasing $T$, and reaches 20% at $T = 1.5$ K. At that temperature the magnetoresistance for the layers with $x = 0.02$ and 0.028 saturates at fields exceeding 5 T. This is important, because at high fields the magnetization also saturates (within the experimental error of our SQUID and MCD experiments). Since the anomalous Hall effect makes a negligible contribution to the slope of $\rho_{Hall}$ (see below), under such conditions we can estimate the hole concentration from the high-field slope of the Hall resistance [1]. The lower panel of Fig. 2(a) illustrates typical Hall data for the $In_{0.98}Mn_{0.02}Sb$ sample. Similar hole concentrations of $p_{Hall} = 2.1 \times 10^{20}$ cm$^{-3}$ were determined in this manner for the layers with $x = 0.02$ and 0.028. Those results are in reasonable agreement with free hole concentrations derived from the c-RBS/PIXE analysis: $p_{RBS/PIXE} = N(Mn_{III}) - 2N(Mn_I) = 2.3 \times 10^{20}$ and $2.7 \times 10^{20}$ cm$^{-3}$ for the samples with $x=0.02$ and 0.028, respectively.

It is well known that the Hall resistivity in a ferromagnetic material can be expressed as the sum of two contributions:

$$\rho_{Hall} = R_0 B + R_A M, \qquad (1)$$

where $R_0$ and $R_A$ are the ordinary and anomalous Hall coefficients, respectively, $B$ is the applied magnetic field, and $M$ is the magnetization [1,19]. Observation of the anomalous Hall effect (AHE) with clear hysteresis loops can decisively establish the ferromagnetism of any new material, whereas SQUID measurements alone are less definitive because ferromagnetic precipitates (*e.g.*, MnSb) may produce similar signature features. As can be seen in the lower panel of Fig 2(a), below $T_\rho$ the Hall resistance of the $In_{0.98}Mn_{0.02}Sb/InSb/CdTe$ sample displays a clear AHE contribution that increases strongly with decreasing *T*. Figure 2(b) shows $\rho_{Hall}$ vs. *B* for four different $In_{1-x}Mn_xSb/InSb/CdTe$ layers at *T* = 1.45 K. Note that both the amplitude of the AHE and the width of the hysteresis loops increase with increasing Mn concentration. The sign of the AHE coefficient $R_A$ is negative (opposite from the normal Hall coefficient in a nonmagnetic *p*-type layer) for all of the $In_{1-x}Mn_xSb$ samples studied, independent of the Mn and hole concentrations. $Ga_{1-x}Mn_xSb$ also displays a negative $R_A$ coefficient [20], whereas $R_A$ is consistently positive in $Ga_{1-x}Mn_xAs$ [1] and in most $In_{1-x}Mn_xAs$ samples [21,22].

These findings are not yet fully understood theoretically. Jungwirth *et al.* recently proposed an approach that relates AHE in III-V ferromagnets to the Berry phase acquired by a quasiparticle wave function upon traversing closed paths on the spin-split Fermi surface [19]. Although $In_{1-x}Mn_xSb$ was not specifically considered, that work derived both positive *and* negative $R_A$ for other III-V ferromagnets, depending on the assumed band parameters and hole concentrations. Since the parameters for $In_{1-x}Mn_xSb$ differ considerably from those of wider-gap III-Mn-Vs, our AHE data provide a further means for sensitively testing the predictive capability of current and future theories.

The ferromagnetic state in $In_{1-x}Mn_xSb$ was further probed by magnetooptical methods. Since our magnetic layers were grown on CdTe/GaAs substrates that are transparent well into the near-IR, we could study not only the magnetooptical Kerr effect in reflection geometry, but also magnetic circular dichroism in transmission. Both experiments employed a lock-in technique and photoelastic modulation. The optical power density from an $Ar^{2+}$ pumped titanium sapphire laser was limited to 5 mW/cm$^2$ to avoid sample heating, and also to minimize the potential for light-induced changes of the magnetization. At $T < T_C$, clear hysteresis loops were observed in both the MCD and MOKE data (the latter will be discussed elsewhere). For the $In_{0.072}Mn_{0.028}Sb$ sample and $\lambda = 900$ nm, Fig. 3(a) shows MCD results collected at various temperatures. The figure plots the field dependence of the MCD signal defined as $(I^+-I^-)/(I^++I^-)$, where $I^+$ and $I^-$ are the transmitted intensities for $\sigma^+$ and $\sigma^-$ circularly-polarized light. The data may be compared with the plot in Fig. 3(b) of the Hall resistivity $\rho_{Hall}(B,T)$ *measured simultaneously* with the MCD signal. The MCD and AHE data are seen to display quite similar temperature-dependent hysteresis loops, which disappear at the same temperature $T_C$. However, the widths of the two sets of loops differ slightly. Since the *simultaneity* of the two measurements eliminates any possibility that the optical excitation plays a role, its cause must lie elsewhere.

Finally, we performed direct SQUID magnetization measurements on the new magnetic alloys. SQUID results for the $In_{0.972}Mn_{0.028}Sb$ sample are illustrated in Fig. 4, which plots *M* vs. *B* at several temperatures, and with *B* applied either perpendicular or parallel to the layer plane. As was the case for the AHE, MCD, and MOKE data, distinct hysteresis loops are observed. At $T = 2$ K and *B* applied in the layer plane, the magnetization is seen to saturate at higher fields, indicating that the easy magnetization axis is perpendicular to the plane (as expected for small tensile strain in the magnetic layer). For a perpendicular field of 10 G, the inset in Fig. 4 shows temperature-dependent remanent magnetizations for two $In_{1-x}Mn_xSb$ layers ($x = 0.02$ and 0.028).

The Curie temperatures determined from these plots, $T_C = 7.0 \pm 0.5$ K and $8.5 \pm 0.5$ K, agree well with the respective $T_\rho$ values, 7.5 K and 9.2 K.

Having probed the magnetic properties of $In_{1-x}Mn_xSb$ in some detail, it is important to determine whether the available models for carrier-mediated ferromagnetism correctly describe this newest and narrowest-energy-gap member of the III-Mn-V family. To our knowledge, only one previous work theoretically treated the ferromagnetic properties of $In_{1-x}Mn_xSb$ and predicted its Curie temperature [6]. Starting from the simplest mean-field theory, that model estimated the $T_C$ enhancement due to exchange and correlation in the itinerant-hole system and the $T_C$ suppression due to collective fluctuations of the ordered moment. Reference [6] concluded that for $In_{1-x}Mn_xSb$ those two effects are small individually, and since they cancel the net influence on $T_C$ is negligible.

We have calculated the Curie temperature in $In_{1-x}Mn_xSb$ using a mean-field-theoretical (MFT) approach based on the 8-band effective band-orbital method (EBOM) [23]. The EBOM is parameterized to yield the correct energy gaps and effective masses at the $\Gamma$ point and energies at the X point, as taken from a recent comprehensive review of III-V band parameters [24]. For narrow-gap materials this model should be more accurate than multiband $k \cdot p$ methods, even if more than six bands [2,6,25] are included, since it produces more physically-correct dispersions at large $k$. We find that the complicated, non-parabolic valence band structure of $In_{1-x}Mn_xSb$ significantly deviates from any simple formula based on a single density-of-states effective mass. The theoretical curves in Fig. 5 plot the Curie temperature as a function of ferromagnetically-active Mn concentration for three values of the free hole concentration: $p = 1 \times 10^{20}$, $2.1 \times 10^{20}$ and $3 \times 10^{20}$ cm$^{-3}$ (recall that $p = 2.1 \times 10^{20}$ cm$^{-3}$ was determined from the transport measurements discussed above). Experimental $T_C$ for the two samples grown with $T_{Mn} = 710$ °C and 720 °C are indicated by the circles and rectangles, respectively. The *x*-axis positions for the open symbols correspond to the *total* Mn concentrations ($x$ = 0.02 and 0.028), whereas the filled symbols plot the ferromagnetically-active *effective* Mn concentrations that account for the $Mn_{III}$-$Mn_I$ anti-ferromagnetic pairing ($x_{eff}$ = 0.016 and 0.021). Clearly the agreement between experiment and theory is much better when the effective Mn concentrations are used. The calculations then yield $T_C$ = 7.7 K and 8.2 K for the two samples, as compared to the experimental SQUID results of 7.0 ± 0.5 K and 8.5 ± 0.5 K.

**4. Conclusions**

We have grown, characterized, and modeled the narrow-gap ferromagnetic semiconductor alloy $In_{1-x}Mn_xSb$. This new material displays all of the attributes of carrier-mediated ferromagnetism, including a well-defined peak in the temperature-dependent resistivity and clear hysteresis loops in the AHE, MCD, MOKE, and SQUID magnetization data. Since $In_{1-x}Mn_xSb$ is the III-Mn-V semiconductor with the largest lattice constant, and some of its band parameters differ considerably from those of all other family members, its measured properties provide a unique test case for further evaluations of the ferromagnetism theories. This material may also offer important advantages in spintronic applications where narrow gap, small mass, or high carrier mobility is desirable.

This work was supported by the DARPA SpinS Program under ONR grant N00014-00-1-0951; by the Center of Excellence CELDIS established under EU Contract No. ICA1-CT-2000-70018 (Poland), by the Director, Office of Science, Office of Basic Energy Sciences, Division of Materials Sciences and Engineering, of the U.S. Department of Energy under Contract No. DE-AC03-76SF00098; and by NSF Grant DMR02-45227.

FIGURE CAPTIONS

Fig. 1. (a) Temperature dependence of zero field resistivity $\rho$ for $In_{1-x}Mn_xSb$/InSb/CdTe ferromagnetic layers with various $x$ as determined by the temperature of the Mn effusion cell (indicated in the figure by $T_{Mn}$). (b) Resistivity *vs.* temperature for $In_{1-x-y}Mn_xBe_ySb$/InSb/CdTe layers grown with constant $T_{Mn}$ = 705 ºC, and with various Be co-doping levels as determined by the temperature of the Be effusion cell (indicated in the figure by $T_{Be}$). The data for a non-magnetic $In_{1-y}Be_ySb$/InSb/CdTe layer with similar hole concentration is also shown for comparison (open circles). In order to avoid extending the vertical scale of the figure, the $\rho$ of the non-magnetic layer is shifted up by 0.08 mΩcm.

Fig. 2. (a) Magnetoresistance $\rho(B)-\rho(0)$ (upper panel) and field dependence of the Hall resistivity $\rho_{Hall}$ (lower panel) measured at various temperatures in the $In_{0.98}Mn_{0.02}Sb$ layer with field applied perpendicular to the layer plane. (b) Hall resistivity $\rho_{Hall}$ *vs.* magnetic field at $T$ = 1.45 K for four $In_{1-x}Mn_xSb$/InSb/CdTe layers. Data for the different samples are shifted along the y-axis; the order and symbols used are the same as in Fig. 1(a).

Fig. 3. Field dependence of: (a) magnetic circular dichroism (MCD) signal and (b) Hall resistivity $\rho_{Hall}$ for the $In_{0.972}Mn_{0.028}Sb$ sample at various temperatures. Both measurements were carried out simultaneously.

Fig. 4. Field dependence of the magnetization measured by SQUID in $In_{0.972}Mn_{0.028}Sb$ at various temperatures, with the field applied either perpendicular or parallel to the layer plane. The inset shows the temperature dependence of the magnetization, measured in a perpendicular field of 10 G, for the two samples with $x$ = 0.02 and 0.028.

Fig. 5. Curie temperature *vs.* Mn concentration for $In_{1-x}Mn_xSb$ epilayers. The curves were calculated by the EBOM-MFT formalism, assuming free hole concentrations: $p = 1\times10^{20}$, $2.1\times10^{20}$ and $3\times10^{20}$ cm$^{-3}$. Experimental $T_C$ for samples grown with $T_{Mn} = 710$ and 720 ºC are indicated by the circles and rectangles, respectively. Open symbols have *x*-axis positions corresponding to the total Mn concentration, while filled symbols correspond to the more relevant ferromagnetically-active concentration (taking into account the $Mn_{III}$-$Mn_I$ anti-ferromagnetic pairing).

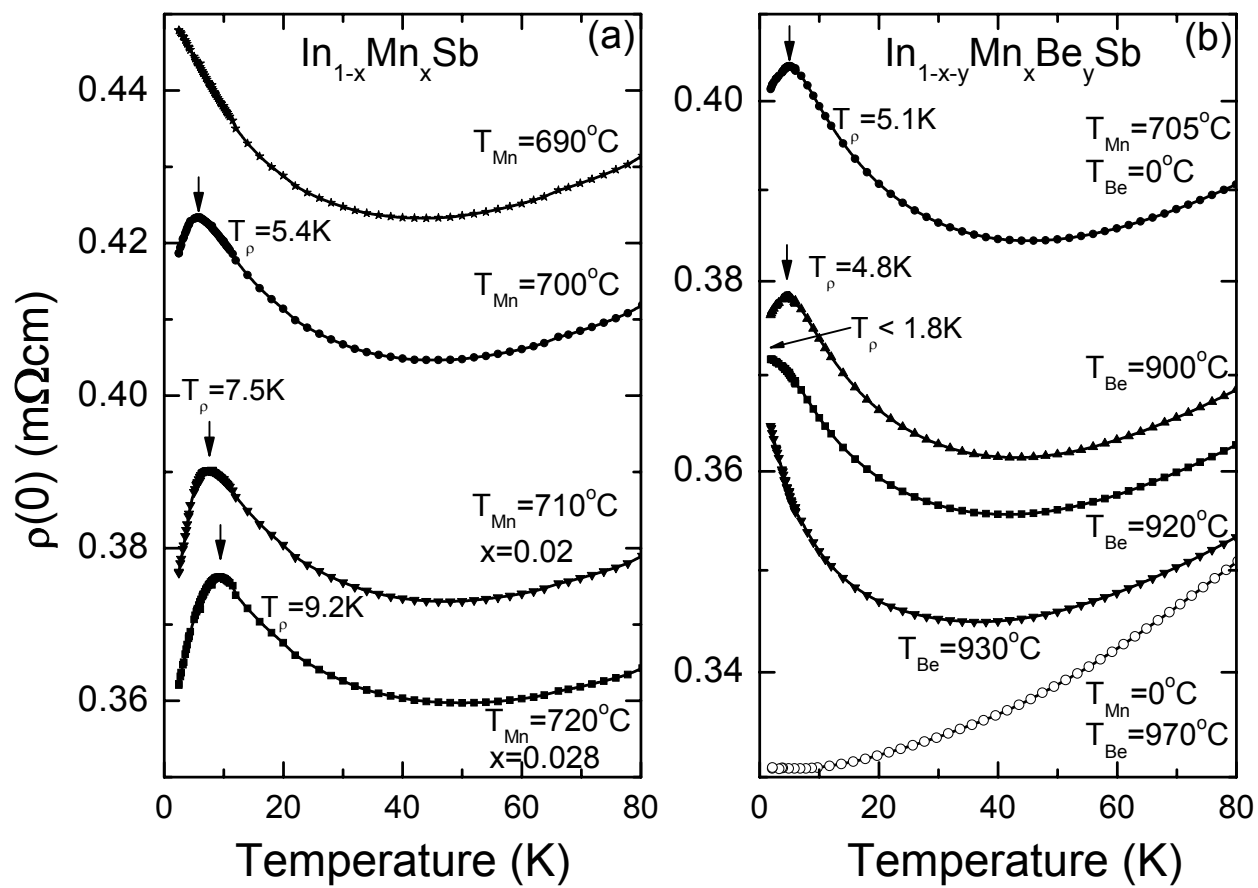

Fig. 1.

Wojtowicz *et al.*

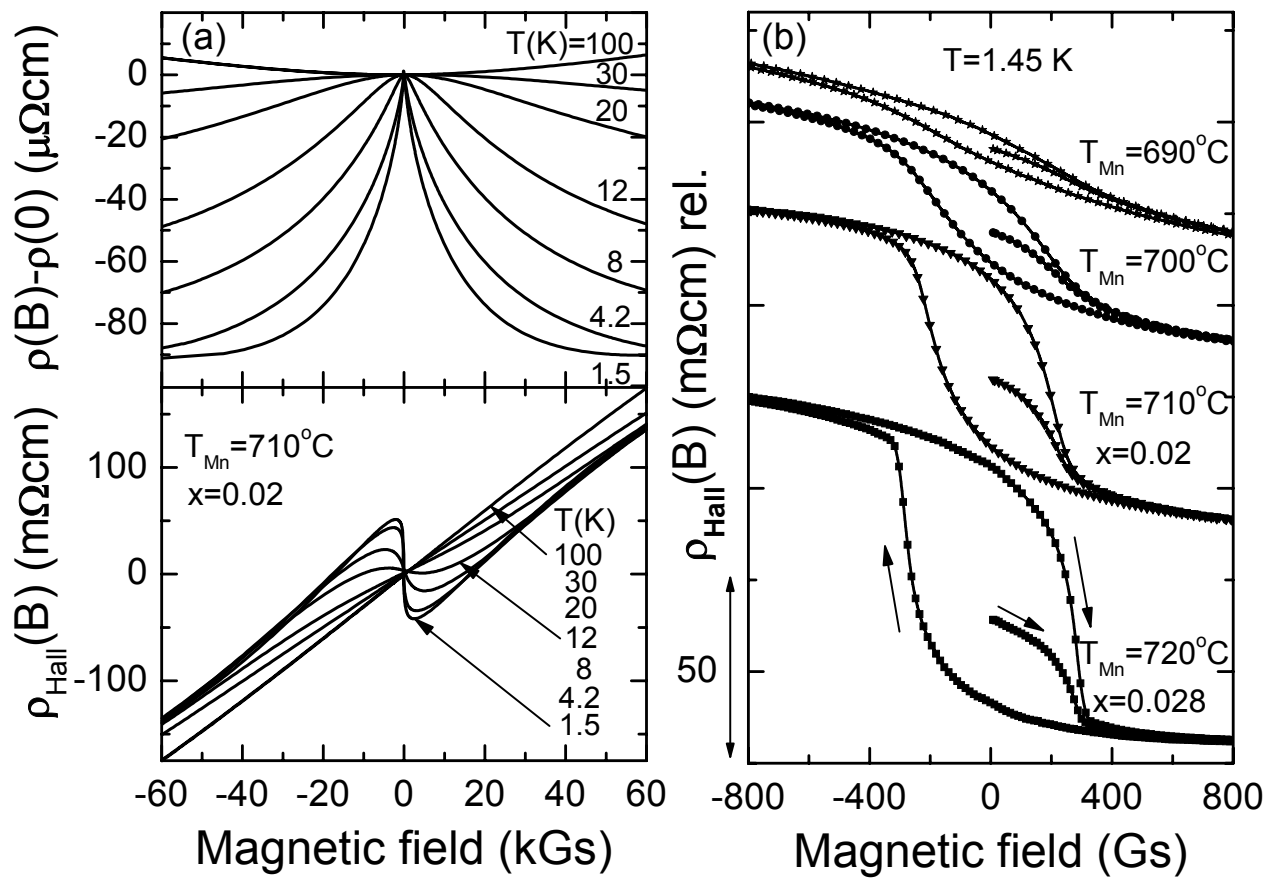

Fig. 2.

Wojtowicz *et al.*

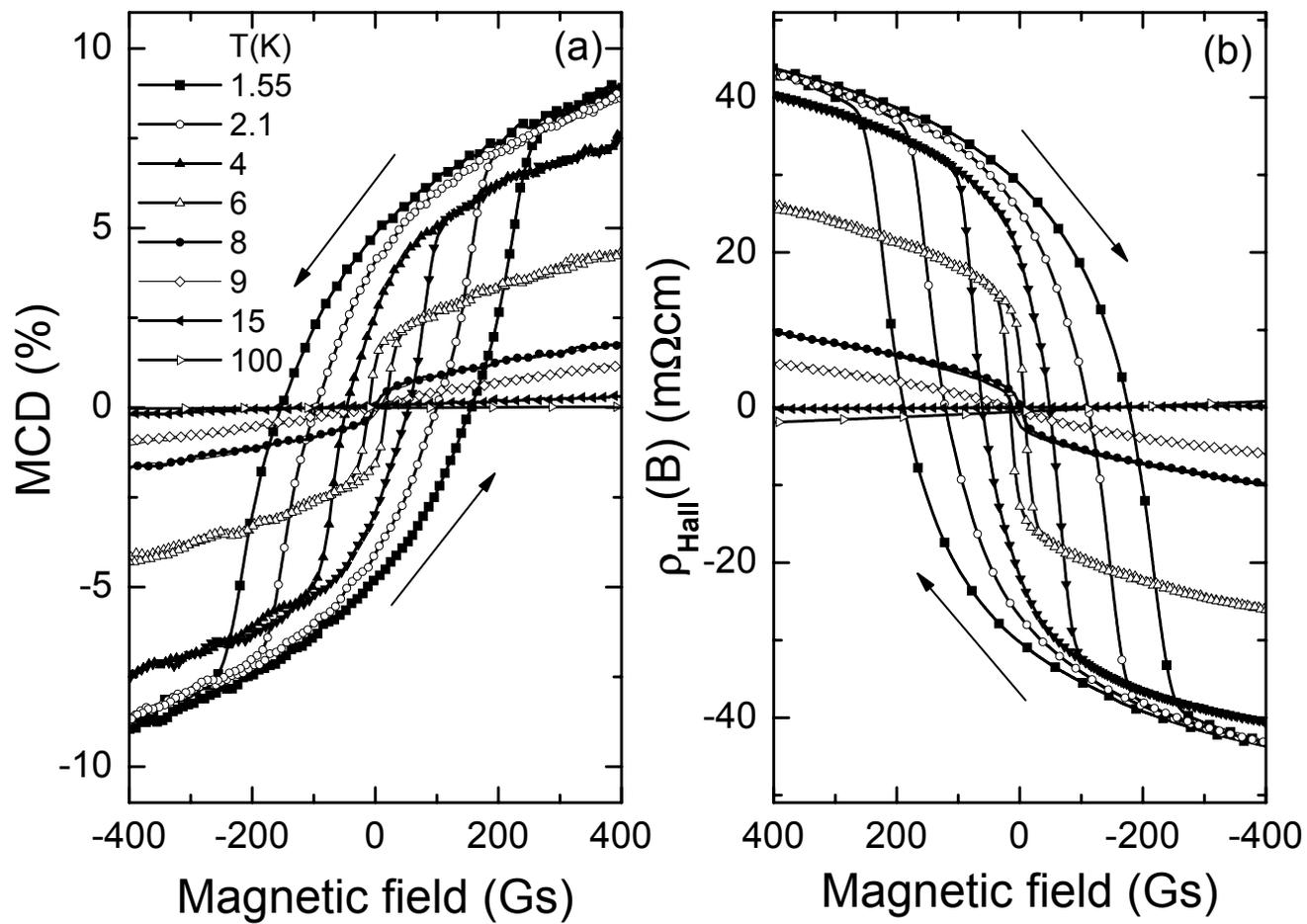

Fig. 3.

Wojtowicz *et al*.

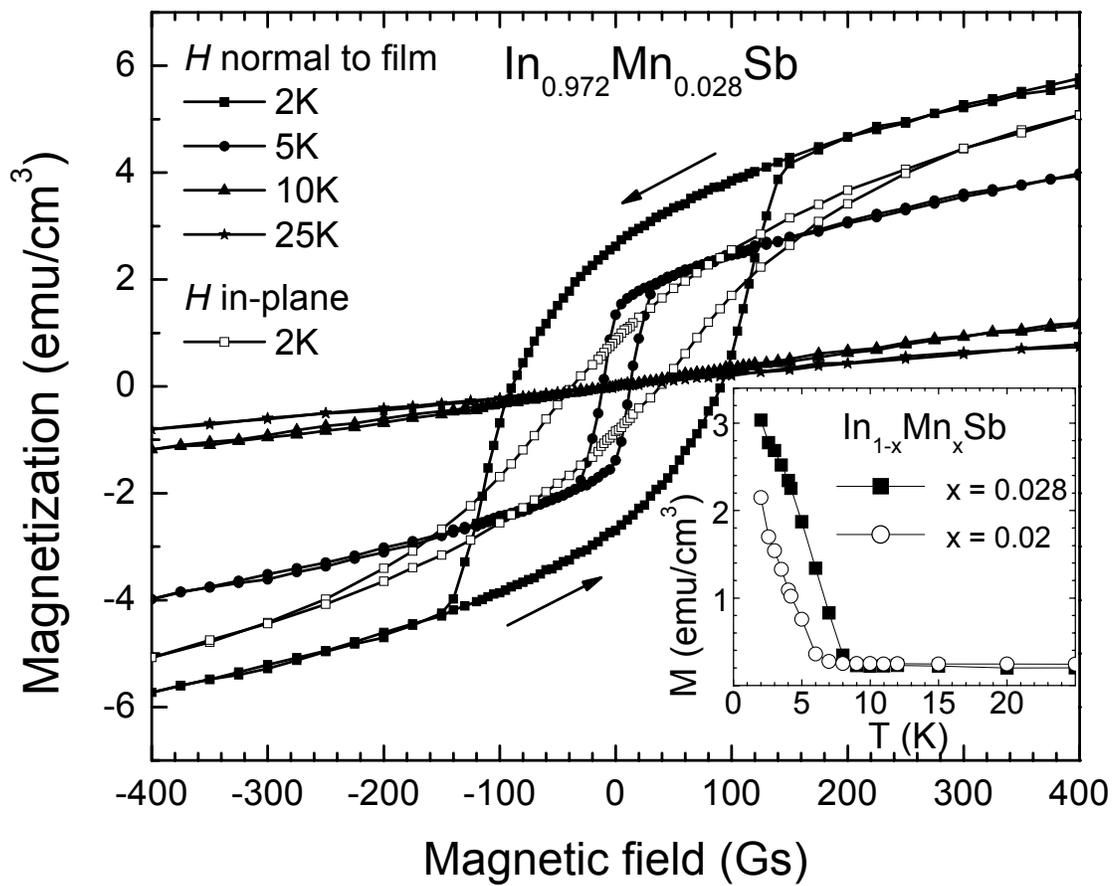

Fig. 4.

Wojtowicz *et al*.

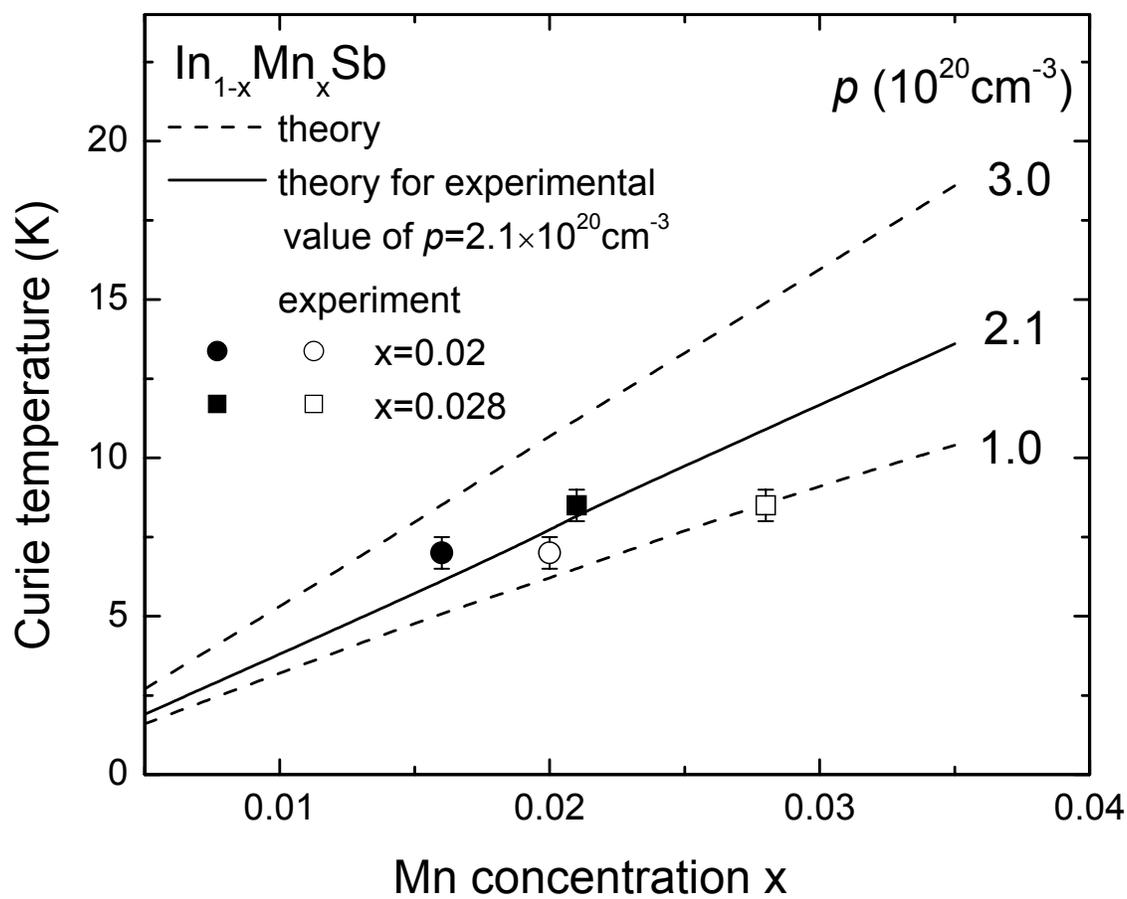

Fig. 5.

Wojtowicz *et al.*